\documentclass[aps,prd,preprintnumbers,superscriptaddress,nofootinbib,showpacs,twocolumn]{revtex4-1}%
\usepackage[dvipdfmx]{graphicx}
\usepackage{bm,latexsym,amsmath,amssymb,amsfonts,mathrsfs}
\usepackage{ulem}  
\usepackage{dcolumn}   
\usepackage[dvipdfmx]{color}

\newcommand*{\MG}{M_{\rm G}}
\newcommand{\simgt}{\lower.5ex\hbox{$\; \buildrel > \over \sim \;$}}
\newcommand{\simlt}{\lower.5ex\hbox{$\; \buildrel < \over \sim \;$}}
\newcommand{\solM}{M_{\odot}}

\begin{document}

\title{Detectability of bigravity with graviton oscillations using gravitational wave observations}


\author{Tatsuya~Narikawa}
\email[Email: ]{narikawa@vega.ess.sci.osaka-u.ac.jp}

\author{Koh~Ueno}
\email[Email: ]{ueno@vega.ess.sci.osaka-u.ac.jp}

\author{Hideyuki~Tagoshi}
\email[Email: ]{tagoshi@vega.ess.sci.osaka-u.ac.jp}
\affiliation{
 Department of Earth and Space Science, Graduate School of Science,
 Osaka University, Toyonaka, Osaka 560-0043, Japan
}

\author{Takahiro~Tanaka}
\email[Email: ]{tanaka@yukawa.kyoto-u.ac.jp}
\affiliation{Department of Physics, Kyoto University, Kyoto 606-8502, Japan }
\affiliation{Yukawa Institute for Theoretical Physics, Kyoto University, 606-8502, Kyoto, Japan}

\author{Nobuyuki~Kanda}
\email[Email: ]{kanda@sci.osaka-cu.ac.jp}
\affiliation{
  Graduate School of Science, Osaka City University, 
  Sumiyoshi-ku, Osaka 558-8585, Japan}

\author{Takashi~Nakamura$^2$}
\email[Email: ]{takashi@tap.scphys.kyoto-u.ac.jp}

\begin{abstract} 
The gravitational waveforms in the ghost-free bigravity theory 
exhibit deviations from those in general relativity.
The main difference is caused by graviton oscillations in the bigravity theory.
We investigate the prospects for the detection of the 
corrections to gravitational waveforms from coalescing compact binaries 
due to graviton oscillations 
and for constraining bigravity parameters with the gravitational wave observations.
We consider the bigravity model discussed by
the De Felice-Nakamura-Tanaka subset of the bigravity model, 
and the phenomenological model in which the bigravity parameters are treated as independent 
variables. 
In both models,
the bigravity waveform shows strong amplitude modulation, 
and there can be a characteristic frequency 
of the largest peak of the amplitude, which depends on the bigravity parameters.
We show that there is a detectable region of the bigravity parameters
for the advanced ground-based laser interferometers, 
such as Advanced LIGO, Advanced Virgo, and KAGRA.
This region corresponds to the effective graviton mass of 
$\mu\simgt10^{-17}~{\rm cm}^{-1}$ for $\tilde{c}-1\simgt10^{-19}$ 
in the phenomenological model, while $\mu\simgt10^{-16.5}~{\rm cm}^{-1}$ 
for $\kappa\xi_c^2 \simgt10^{0.5}$ in the De Felice-Nakamura-Tanaka subset of the bigravity model, respectively,
where $\tilde{c}$ is the propagation speed of the massive graviton
and $\kappa\xi_c^2$ corresponds to the corrections to the gravitational constant in general relativity.
These regions are not excluded by existing solar system tests.
We also show that, 
in the case of $1.4-1.4\solM$ binaries at the distance of $200~{\rm Mpc}$,
$\log\mu^2$ is determined with an accuracy of ${\cal O}$(0.1)\% at the 1$\sigma$ level 
for a fiducial model with $\mu^2=10^{-33}~{\rm cm}^{-2}$
in the case of the phenomenological model.
\end{abstract}

\preprint{OUTAP-365, KUNS-2549, YITP-15-20}
\date{\today}
\maketitle

\section{Introduction}
The second-generation laser interferometers such as 
Advanced LIGO~\cite{Harry:2010zz}, Advanced Virgo~\cite{Virgo}, and KAGRA~\cite{Somiya:2011np,Aso:2013eba},  
will be in full operation within a few years.
These detectors are sensitive to gravitational waves (GWs) in the frequency band between $10$ 
and $\sim1000~{\rm Hz}$.
The inspiral of a coalescing compact binary (CCB) system
is one of the most promising sources for these detectors.
These detectors will be able to see CCB systems, composed of neutron stars and/or stellar-mass black holes (BHs), 
within $200-1000~{\rm Mpc}$. 
GW observations of the inspiral signals from CCB systems can be a powerful 
tool to probe strong-field, 
dynamical aspects of gravity theories~\cite{Yunes:2013dva}.
One of the science targets of these projects is to test the correctness of 
general relativity (GR) through comparison of observed gravitational waveforms with the prediction.

Cosmological observations of distant type Ia supernovae have discovered 
the late-time accelerated expansion of the Universe~\cite{Riess:1998cb, Perlmutter:1998np}.
Observations of the type Ia supernovae, the cosmic microwave background anisotropies, 
and the large scale structure of galaxies consistently suggest the current cosmic acceleration.
However, the origin of this late-time cosmic acceleration is still unknown, and it is one of the biggest 
unsolved problems in cosmology. 
It may suggest the existence of dark energy.
But it may also suggest a sign of breakdown of GR on cosmological scales,
and motivates many researchers to study modified gravity (MG) models as cosmological models 
(see, e.g., \cite{Clifton:2011jh} for a review).

As an alternative model to GR,
we focus on the first example of the ghost-free bigravity model\cite{Hassan:2011zd},  
which is constructed based on the fully nonlinear massive gravity theory
~\cite{deRham:2010ik,deRham:2010kj,Hassan:2011hr}
(see, e.g., \cite{Hinterbichler:2011tt,deRham:2014zqa} for a review).
In the ghost-free massive gravity model, it is difficult to construct spatially flat 
Friedmann-Lema\^\i tre-Robertson-Walker (FLRW) solutions
~\cite{D'Amico:2011jj,DeFelice:2012mx,Gumrukcuoglu:2012aa,D'Amico:2012zv},
\footnote{
Recently, other such pathologies on superluminal propagation, problems of acausality,
have been argued in \cite{Izumi:2013poa,Deser:2013eua}.}
while in the ghost-free bigravity model spatially flat FLRW solutions exist
~\cite{Comelli:2011zm}.
The bigravity model with a small mass is interesting phenomenologically.
However, such models do not remain to have a healthy background cosmological 
solution at a high energy. Therefore, it would be required to embed the model into 
a more fundamental theory that is valid even at higher energies.
The first attempt was made in Ref.~\cite{Yamashita:2014cra}, in which the bigravity
model is shown to be embedded in the Dvali-Gabadadze-Porrati 2-brane model~\cite{Dvali:2000hr},
at least at low energies.

In the bigravity theory GWs propagate differently from those in GR.
So direct GW observations will be a powerful probe of the bigravity theory.
In the ghost-free bigravity theory, physical and hidden modes of GWs are both excited.
These two gravitons interfere with each other like neutrinos during their propagation, 
which is called the graviton oscillation and
the observed GWs exhibit deviations from GR~\cite{DeFelice:2013nba}.
While there are previous studies on the modified propagation of gravitational 
waves due to the finite mass of the graviton 
(see e.g., Refs.~\cite{Will:1997bb,Finn:2001qi,Yagi:2009zm,Hazboun:2013pea}), 
those studies were based on the linearized Fierz-Pauli theory~\cite{Fierz:1939ix} and
did not care about the appearance of a ghost mode at the nonlinear level.
Once we care about the ghost appearance, we need to consider the ghost-free
massive gravity, but the simplest model does not have a suitable FLRW background 
solution, as we mentioned earlier. In the case of bigravity, the situation is very different 
since we have two gravitons. Furthermore, the linear theory is not sufficient to discuss 
the solar system constraint, and the generation and propagation of GWs in this model.
Owing to the Vainshtein mechanism, the ghost-free bigravity model can give almost the same
prediction as GR at least in the weak-field case. However, the gravitational waveforms differ from 
those in GR, because of the graviton oscillation effect.
De Felice, Nakamura and Tanaka~\cite{DeFelice:2013nba} (DFNT)
have pointed out that the interesting parameter range of 
graviton mass exists, where large deviations from the GR case are produced  in GW signals,
while it cannot be excluded by the solar system tests.
So, one can use gravitational waveforms to identify the effect of modified gravity.

To evaluate the parameter estimation accuracy,
the Fisher matrix has often been used~\cite{Cutler:1994ys,Poisson:1995ef}.
Many works~\cite{Will:1997bb,Yagi:2009zm,Barvinsky:2002gg, Berezhiani:2007zf} 
have been done to study the possibility to test the modified propagation 
of GWs due to the graviton mass by using the Fisher matrix.
Bayesian hypothesis testing is also useful for model selection in the GW data analysis~\cite{DelPozzo:2011pg}.
Recently, Vallisneri~\cite{Vallisneri:2012qq} has introduced 
a simple method to test modified gravity within the framework of the Bayesian hypothesis testing. 
In this method, one can compute the odds ratio from the fitting factor between 
the general relativistic and modified gravity's waveforms.
More recently, Del Pozzo {\it et al}.~\cite{DelPozzo:2014cla} have compared the prediction from 
Vallisneri's approximate formula against an exact numerical calculation of the Bayes factor.
They found that the approximate formula recovers the numerical result with good accuracy.

In this paper, 
we explore the detectability of the bigravity corrections due to the graviton oscillation
to the waveforms from CCBs.
We consider nonspinning binary systems consisting of binary neutron stars (BNS) with $1.4-1.4\solM$,
as well as neutron star$-$black hole binaries (NSBH) with $1.4-10\solM$ and 
binary black holes (BBH) with $10-10\solM$.
We consider two kinds of bigravity models; 
one is the phenomenological model that is constructed phenomenologically to investigate the possibility 
to constrain the bigravity parameters only from GW observation,
and the other is the DFNT subset of the bigravity model in which the bigravity parameters are set to give 
a consistent cosmological model.
We examine the detectability in both models by using Vallisneri's formulas 
assuming the observations with the advanced ground-based laser interferometers.
We also evaluate the measurement accuracy of bigravity parameters
by using the Fisher matrix.
We assume the noise power spectrum density of advanced LIGO that is called 
Zero Det, High Power~\cite{AdvLIGOZDHP}.
We take the lowest frequency to be $f_{\rm low}=$20Hz. 

The remainder of this paper is organized as follows.
In Sec.~\ref{sec:bigravity}, we review the ghost-free bigravity model,
and the derivation of the modified waveforms.
In Sec.~\ref{sec:statistics}, we briefly review the Vallisneri's formulas to evaluate the detectability
of the bigravity model
and the Fisher matrix to evaluate
the measurement accuracy of the bigravity parameters.
In Sec.~\ref{sec:phenom}, we 
show the detectable region of the bigravity model in the phenomenological model
on the model parameter space. 
We discuss the physical explanation on how 
the detectable range is determined, and 
the correspondence of the detectable range with 
the fitting factor between the GR and bigravity waveforms.
We also evaluate the measurement accuracy of 
the bigravity parameters. 
In Sec \ref{sec:DFNT}, we discuss the detectability of the bigravity corrections to gravitational waveforms 
in the DFNT subset.
Section~\ref{sec:summary} is devoted to summary and conclusions.
%

\section{Gravitational waves in the bigravity model}
\label{sec:bigravity}
In this section, we briefly review the graviton oscillations in the ghost-free bigravity model.
\subsection{Ghost-free bigravity theory}
We describe the first example of ghost-free bigravity model~\cite{Hassan:2011zd}.  
The action of this model is given as
\begin{eqnarray}
  S & = & \frac{\MG^2}{2} \int d^{4} x\sqrt{- {\rm det}~g} R[g]
  + \frac{\kappa\MG^{2}}{2} \int d^{4}x\sqrt{- {\rm det}~\tilde{g}} \tilde{R}[\tilde{g}] \nonumber\\
  & & -m^{2}\MG^2 \int d^{4} x\sqrt{- {\rm det}~g} \sum_{n=0}^{4} c_{n} V_{n} (Y_{\nu}^{\mu}) 
 +S_{\rm m}[g] \,, 
 \label{Lagrangian}
\end{eqnarray}
where $Y_{\nu}^{\mu}=\sqrt{g^{\mu\alpha}\tilde{g}_{\alpha\nu}}$, and
$V_n$ are elementary symmetric polynomials~\cite{deRham:2010kj} defined as
\begin{eqnarray}
V_{0} & = & 1~,\quad V_{1}=[Y],\quad V_{2}=[Y]^{2}-[Y^{2}]~,\nonumber\\
V_{3} & = & [Y]^{3}-3[Y][Y^{2}]+2[Y^{3}]~,\nonumber\\
V_{4} & = & [Y]^{4}-6[Y]^{2}[Y^{2}]+8[Y][Y^{3}]+3[Y^{2}]^{2}-6[Y^{4}],~\nonumber\\
\end{eqnarray}
where the trace of $Y^{n}$ is expressed as $[Y^{n}]={\rm tr}(Y^{n})
=Y_{\alpha_{1}}^{\alpha_{0}}Y_{\alpha_{2}}^{\alpha_{1}}\cdots Y_{\alpha_{0}}^{\alpha_{n-1}}$.
$c_{n}$ are dimensionless constants
and the matter action $S_{{\rm m}}[g]$ only couples to the physical metric $g_{\mu\nu}$.
$\tilde{g}_{\mu\nu}$ is an additional dynamical tensor field, which we refer to as the hidden metric.
$R$ and $\tilde{R}$ denote the scalar curvatures for $g_{\mu\nu}$ and $\tilde{g}_{\mu\nu}$, respectively.
$\MG=1/(8\pi G_{\rm N})$ is the Planck mass,
$\kappa$ is a constant that expresses the ratio between the two gravitational constants 
for $\tilde{g}_{\mu\nu}$ and $g_{\mu\nu}$ 
 and graviton mass parameter $m^2$ 
can be absorbed into the parameters $c_n$.
The action consists of the standard Einstein-Hilbert kinetic terms for both $g_{\mu\nu}$ and 
$\tilde{g}_{\mu\nu}$,
and coupling terms between  $g_{\mu\nu}$ and $\tilde{g}_{\mu\nu}$.
The theory 
is free from the Boulware-Deser ghost~\cite{Boulware:1973my} 
in both $g_{\mu\nu}$ and $\tilde{g}_{\mu\nu}$ sectors~\cite{deRham:2010ik,deRham:2010kj,Hassan:2011hr}.

We assume the spatially flat FLRW background~\cite{Comelli:2011zm}
\begin{eqnarray}
  ds^{2}=a^{2}(- c^2 dt^{2}+d\bm{x}^{2})\,,\quad d\tilde{s}^{2}=\tilde{a}^{2}(-\tilde{c}^{2}dt^{2}+d\bm{x}^{2})\,,
\end{eqnarray}
where the scale factors $a$, $\tilde{a}$ and the propagation speed of the hidden graviton $\tilde{c}$ are 
functions of the conformal time coordinate $t$. (Hereafter, we set $c=1$.)
We focus on a healthy branch of background cosmological solutions~\cite{Comelli:2012db,DeFelice:2013nba}, in which 
$\tilde{c}\tilde{a}\dot{a}-a\dot{\tilde{a}}=0$,
where $\dot{}\equiv d/dt$.
We also focus on the case of $m^{2}\gg\rho_{\rm m}/\MG^{2}$, where $\rho_{\rm m}$ is the matter energy density. 
In this limit, we can regard $\xi\equiv \tilde{a}/a$ as a constant, $\xi_c$.
Now the usual Friedmann equation for the physical metric $g_{\mu\nu}$ is given as
\begin{eqnarray}
 3H^2 \approx \tilde{M}_{\rm G}^{-2}\rho_{\rm m}, 
\end{eqnarray}
where $H\equiv \dot{a}/a^2$ is the Hubble parameter
and $\tilde{M}_{\rm G}^2\equiv \MG^2(1+\kappa\xi_c^2)$ is the effective gravitational constant.

\subsection{Propagation of gravitational waves}
By using the nonlinear Hamiltonian analysis~\cite{Hassan:2011hr},
we find that there are in general seven propagation degrees of freedom in the ghost-free bigravity theory. 
The seven modes consist of one massive and one massless spin-2 fields.
Dominant contributions to GW radiation in the theory are 
two plus two helicity-2 modes for physical and hidden sectors,
both of which are generated in the same way as in GR~\cite{DeFelice:2013nba}.
Here we consider the double FLRW background solutions and denote the perturbations around them as
$\delta g_{ij}=a^{2}(h_{+}\varepsilon_{ij}^{+}+h_{\times}\varepsilon_{ij}^{\times})$
and $\delta \tilde{g}_{ij}=\tilde{a}^{2}(\tilde{h}_{+}\varepsilon_{ij}^{+}+\tilde{h}_{\times}
\varepsilon_{ij}^{\times})$, 
where $\varepsilon_{ij}^{+\times}$ represent the polarization tensors for plus and cross modes.

The physical and hidden gravitational modes mix during their propagation, 
because of their coupling through the interaction term.
The mixing of the gravitational wave modes is interpreted as graviton oscillations  
in analogy with neutrino oscillations.

Neglecting the effects of cosmic expansion, we have the following propagation equations 
for gravitational waves~\cite{Comelli:2012db,DeFelice:2013nba}:
\begin{eqnarray}
 \ddot{h}-\Delta h+m^{2}\Gamma_{c}(h-\tilde{h})&=&0,\nonumber\\
 \ddot{\tilde{h}}-\tilde{c}^{2}\Delta\tilde{h}+\frac{m^{2}\Gamma_{c} \tilde{c}}
 {\kappa\xi_{c}^{2}}(\tilde{h}-h)&=&0\,,\label{GWeqs}
\end{eqnarray}
where $\xi_c$ and $\Gamma_c$ are constants.
Later, $\Gamma_c$ is absorbed into the effective mass for the graviton defined as
$\mu^2\equiv (1+1/\kappa\xi_c^2)m^2\Gamma_c$.
Since the propagation equations are identical for both polarizations, 
we have omitted the index $+/\times$.
Solving Eqs. (\ref{GWeqs}), we obtain two eigen wave numbers for a given gravitational wave frequency $f$ as
\begin{eqnarray}
k_{1,2}^2 = (2\pi f)^2 - \frac{\mu^2}{2}\left( 1+x \mp \sqrt{1 + 2x\frac{1-\kappa\xi_c^2}{1+\kappa\xi_c^2}+x^2} \right), \nonumber\\
\end{eqnarray}
and the corresponding eigenfunctions are given as
\begin{eqnarray}
h_{1} & = & \cos\theta_{g}\, h+\sin\theta_{g}\sqrt{\kappa}\xi_{c}\,\tilde{h},\\
h_{2} & = & -\sin\theta_{g}\, h+\cos\theta_{g}\sqrt{\kappa}\xi_{c}\,\tilde{h},
\end{eqnarray}
with the mixing angle 
\[
\theta_{g}=\frac{1}{2}\cot^{-1}\left(\frac{1+\kappa\xi_{c}^{2}}{2\sqrt{\kappa}\xi_{c}}x
+\frac{1-\kappa\xi_{c}^{2}}{2\sqrt{\kappa}\xi_{c}}\right)~,
\]
and
\begin{eqnarray}
x\equiv\frac{2(2\pi f)^{2}(\tilde{c}-1)}{\mu^{2}}\,. \label{eq:x}
\end{eqnarray}
In the case of the usual Vainshtein mechanism, the Compton wavelength of the graviton 
should be as large as $300~{\rm Mpc}$ or so to pass the solar system constraints.
In that case the effect of the graviton mass is hardly detected even if we consider the propagation 
of GWs over the cosmological distance scale. However, in the bigravity model discussed in Ref~\cite{DeFelice:2013nba},
thanks to the enhanced Vainshtein mechanism, it is possible to keep the effective graviton
mass $\mu$ much larger~\cite{DeFelice:2013nba}.
When the Vainshtein mechanism~\cite{Vainshtein:1972sx} works, metric tensor perturbations 
on both sectors are equally excited inside the Vainshtein radius.

\subsection{Modified inspiral waveforms due to graviton oscillations}
\label{sec:modwave}
Here we discuss only the inspiral phase of gravitational waves from CCB systems
in the ghost-free bigravity model.
Both $h$ and $\tilde{h}$ are excited exactly as in the case of GR
~\cite{DeFelice:2013nba}.
By using the stationary phase approximation,
the observed signal in the frequency domain is given as 
\footnote{For simplicity, we assume a signal from a face-on binary system at the zenith.}
\begin{eqnarray}
 h(f)={\cal A}_{}(f)e^{i\Phi(f)}\left[B_{1}e^{i\delta\Phi_{1}(f)}+B_{2}e^{i\delta\Phi_{2}(f)}\right]~,\label{vf}
\end{eqnarray}
where the amplitude ${\cal A}(f)$ (up to Newtonian order), the bigravity corrections $B_{1,2}$ and
the phase function $\Phi(f)$ (up to 3.5PN order), 
and the phase corrections $\delta\Phi_{1,2}$ are given as
\begin{eqnarray}
{\cal A}(f) & = & \sqrt{\frac{5\pi}{24}}\frac{{\cal M}^{2}}{(8\pi M_{\rm G}^2)^2D_L}y^{-7/6},\\
B_{1} & = & \cos\theta_{g}(\cos\theta_{g}+\sqrt{\kappa}\xi_{c}\sin\theta_{g}),\\
B_{2} & = & \sin\theta_{g}(\sin\theta_{g}-\sqrt{\kappa}\xi_{c}\cos\theta_{g}),\\
\Phi(f) & \equiv & 2\pi ft_{c}-\Phi_{c}-\pi/4+\frac{3}{128}y^{-5/3} \Bigl\{ 1+ \nonumber\\
 &  & \hspace{-1cm}+\left(\frac{3715}{756}+\frac{55}{9}\eta\right)\eta^{-2/5}y^{2/3}
 -16\pi \eta^{-3/5}y \nonumber\\
  &  & \hspace{-1cm} + \left( \frac{15~293~365}{508~032} + \frac{27~145}{504}\eta 
  + \frac{3085}{72}\eta^2 \right)\eta^{-4/5} y^{4/3} \nonumber\\
  &  & \hspace{-1cm} + \left( \frac{38~645}{756} - \frac{65}{9}\eta \right) \left[ 1
  + \ln \left( \frac{y}{y_{\rm ISCO}} \right) \right] \pi \eta^{-1} y^{5/3}  \nonumber \\
  &  & \hspace{-1cm} + \left[ \frac{11~583~231~236~531}{4~694~215~680} 
  - \frac{640}{3} \pi^2 - \frac{6848}{21} \gamma_{\rm E}  \right. \nonumber \\
  &  & \hspace{-1cm} - \frac{6848}{63} \ln(64\eta^{-3/5}y) + \left( 
  -\frac{15~737~765~635}{3~048~192} \right.  \nonumber \\
  &  & \hspace{-1cm} \left.\left. + \frac{2255}{12}\pi^2 \right)\eta + \frac{76~055}{1728} \eta^2 
  - \frac{127~825}{1296} \eta^3 \right] 
  \eta^{-6/5} y^{2} \nonumber \\
    &  & \hspace{-1cm} + \left( \frac{77~096~675}{254~016} + \frac{378~515}{1512} \eta 
    - \frac{74~045}{756} \eta^2 \right) \pi \eta^{-7/5} y^{7/3}
    \Bigl\} \,, \nonumber \\ \\
\delta\Phi_{1,2} & = & -\frac{\mu D_L\sqrt{\tilde{c}-1}}{2\sqrt{2x}}\left(1+x\mp
\sqrt{1+x^{2}+2x\frac{1-\kappa\xi_{c}^{2}}{1+\kappa\xi_{c}^{2}}}\right), \nonumber\\ \label{deltaPhi}
\end{eqnarray}
where $y \equiv {\cal M} f / (8 \tilde{M}_{\rm G}^2)$, 
${\cal M}\equiv(m_{1}m_{2})^{3/5}/(m_{1}+m_{2})^{1/5}$ is the chirp mass,
$\eta={m_{1}m_{2}}/{(m_{1}+m_{2})^{2}}$ is the symmetric mass ratio, 
$t_c$ is the coalescence time and $\Phi_c$ is the phase at the coalescence.
$\gamma_{\rm E}=0.577~216~.~.~.$ is the Euler constant. 
$D_L$ is the luminosity distance to the source.
\footnote{The phase shifts are not integer powers of post-Newtonian (PN) expansion parameter $y$.} 
The first and second terms in Eq.~(\ref{vf}) show the contributions
of $h_{1}$ and $h_{2}$, respectively.
In the above waveform, we can take the following five parameters as independent parameters for GR,
$\theta_{\rm GR}=\{D_L, m_1, ~m_2, ~t_c, ~\Phi_c \}$. 
On the other hand, there are eight independent parameters for the phenomenological bigravity model,
$\theta_{\rm MG}=\{\log\mu^2, ~\log(\tilde{c}-1),~\kappa\xi_c^2, ~\theta_{\rm GR} \}$. 
In the DFNT subset of the bigravity model, $\log(\tilde{c}-1)$ is not an independent variable, 
but it depends on the matter density.

\begin{figure}[thbp]
  \begin{center}
   \includegraphics[width=0.50\textwidth,angle=0]{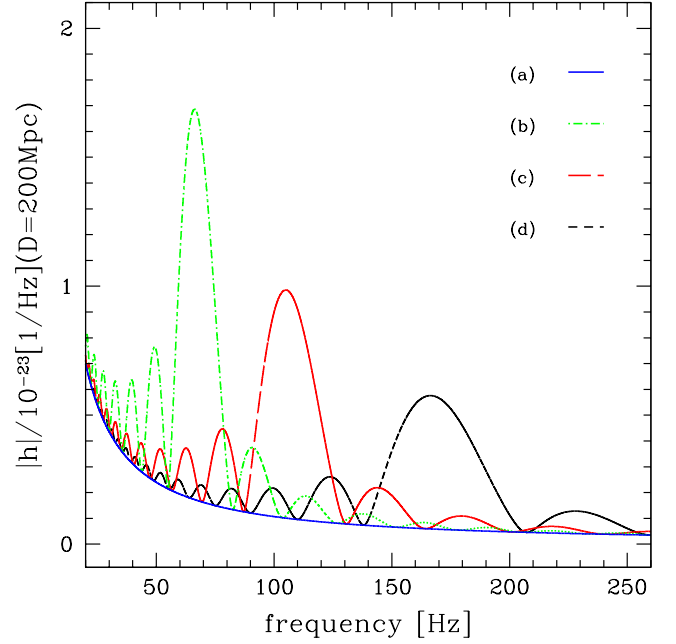}
  \end{center}
  \caption{
The frequency-domain gravitational waves $h(f)$ 
for different values of the model parameter sets of $(\mu^2,~\tilde{c}-1)$.
The curves are plotted for (a) GR [solid (blue)] and for the bigravity models with 
(b) $(\mu^2,~\tilde{c}-1)=(10^{-33.2}~{\rm cm}^{-2},~10^{-17.8})$ [dot-dashed (green)],
(c) $(10^{-33}~{\rm cm}^{-2},~10^{-18})$ [long-dashed (red)], and (d) $(10^{-32.8}~{\rm cm}^{-2},~10^{-18.2})$ 
[dashed (black)], respectively, at fixed $\kappa\xi_c^2=100$.
Here we consider BNS at the distance, $D_L=200~{\rm Mpc}$. 
The SNR and the fitting factor between the GR waveform and  each waveform in this figure 
become as follows: $({\rm SNR}, {\rm FF})=$ (a) $(8.7, 1.0)$, (b) $(31, 0.50)$, (c) $(26, 0.47)$, (d) $(21, 0.53)$.
The definition of FF is given in Eq.~(\ref{FF}).
}%
\label{fig:hFD}
\end{figure}

\begin{figure}[thbp]
  \begin{center}
    \includegraphics[keepaspectratio=true,height=85mm]{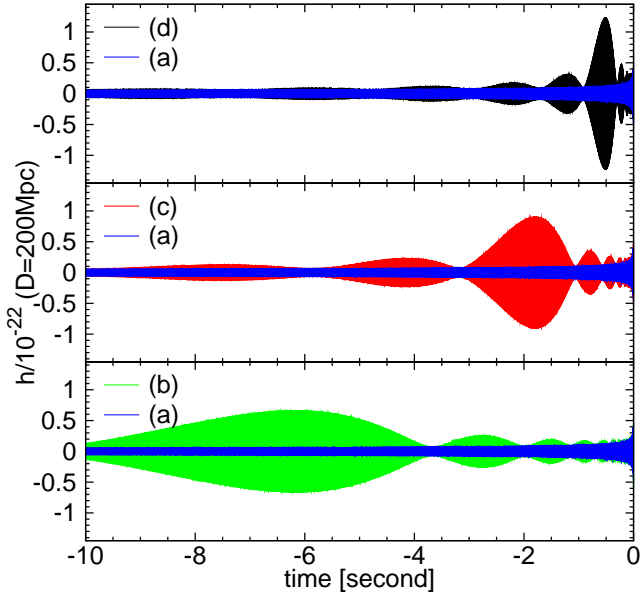}
  \end{center}
  \caption{
The time-domain gravitational waveform $h(t)$. The coalescence time $t_c$ is set to 0.
The parameters and the definitions of the curves are the same as those of Fig.~\ref{fig:hFD}. 
}%
\label{fig:hTD}
\end{figure}

From Eq. (\ref{vf}), we have the formula for the amplitude of the wave in frequency domain.
\begin{eqnarray}
\left| h(f) \right|&=&{\cal A}(f) \left(1+2B_1 B_2 (\cos(\Delta\delta\Phi)-1)\right)^{1/2}, \label{eq:abshf} \\
\Delta \delta\Phi &\equiv& \delta\Phi_1-\delta\Phi_2. \label{eq:DeltaPhi}
\end{eqnarray}
Thus, unless $B_1B_2$ or $\Delta\delta\Phi$ is zero, 
amplitude modulation occurs in the bigravity waveform that is caused by 
the interference between two modes. 
The peak amplitude of the modulated waveform is determined  
by $1+2B_1 B_2 (\cos(\Delta\delta\Phi)-1)$. 

Figure~\ref{fig:hFD} shows the frequency-domain gravitational waveform
for BNS with $1.4-1.4\solM$ and $D_L=200~{\rm Mpc}$.
Curves are for different sets of $(\mu^2,~\tilde{c}-1)$ at fixed $\kappa\xi_c^2=100$.
Figure~\ref{fig:hTD} shows the same gravitational waveforms in the time domain, 
where the coalescence time $t_c$ is set to 0.
Curves in Figs.~\ref{fig:hFD}$-$\ref{fig:hTD} are for
(a) GR [solid (blue)] and for the bigravity with
$(\mu^2,~\tilde{c}-1)=$  
(b) $(10^{-33.2}~{\rm cm}^{-2},~10^{-17.8})$ [dot-dashed (green)], 
(c) $(10^{-33}~{\rm cm}^{-2},~10^{-18})$ [long-dashed (red)], and 
(d) $(10^{-32.8}~{\rm cm}^{-2},~10^{-18.2})$ [dashed (black)], respectively.
We find that the waveforms of the bigravity model are significantly different from those of GR. 
In particular, there is a characteristic largest peak in the modulated waveform.
The frequency at the highest peak amplitude can be explained in the following way.
In Ref.~\cite{DeFelice:2013nba}, De Felice {\it et al}. showed that measurable 
effects are expected only when $x\approx1$.
Using Eq.~(\ref{eq:x}) we can estimate the characteristic frequency corresponding to $x\approx1$:
\begin{eqnarray}
 f_{\rm peak} \equiv \frac{1}{2\pi} \left( \frac{\mu^2}{2(\tilde{c}-1)} \right)^{1/2}.
\label{eq:fpeak}
\end{eqnarray}

The corresponding time at the highest peak is given as
\begin{eqnarray}
 \tau_{\rm peak} \equiv t_c - t_{\rm peak} = \frac{5}{256} \frac{1}{\eta (\pi f_{\rm peak})^{8/3} M_t^{5/3}},
\end{eqnarray}
with the total mass $M_t=m_1+m_2$ .

\begin{figure}[tbp]
  \begin{center}
   \includegraphics[width=0.50\textwidth,angle=0]{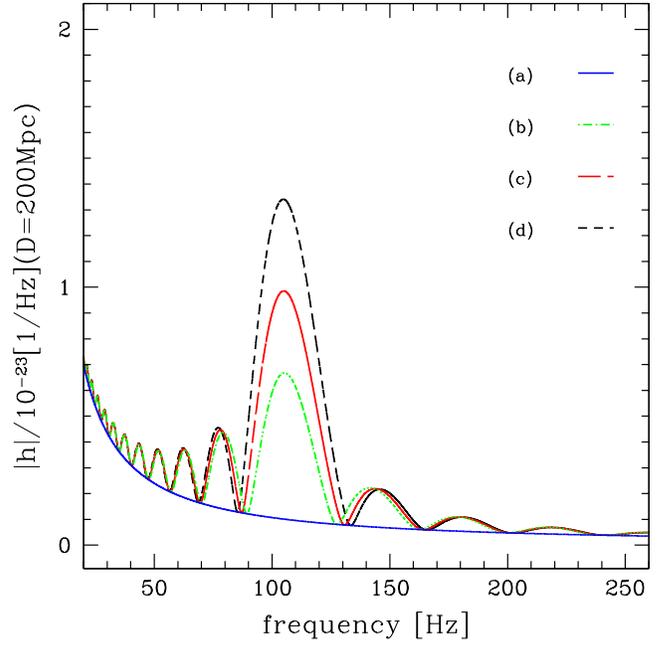}
  \end{center}
  \caption{
The same as Fig.~\ref{fig:hFD} but for different values of $\kappa\xi_c^2$ 
in the case of $(\mu^2,~\tilde{c}-1)=(10^{-33}~{\rm cm}^{-2},~10^{-18})$.
The curves are for (a) GR [solid (blue)] and the bigravity model with (b) $\kappa\xi_c^2=50$ 
[dot-dashed (green)],
(c) $\kappa\xi_c^2=100$ [long-dashed (red)], and (d) $\kappa\xi_c^2=1000$ [dashed (black)], respectively.
Each curve corresponds to $({\rm SNR},~{\rm FF})=$ (a) $(8.7,~1.0)$, (b) $(19,~0.58)$, 
(c) $(26,~0.47)$, (d) $(34,~0.41)$.
}%
\label{fig:hFD_kchange}
\end{figure}

\begin{figure}[tbp]
  \begin{center}
    \includegraphics[keepaspectratio=true,height=85mm]{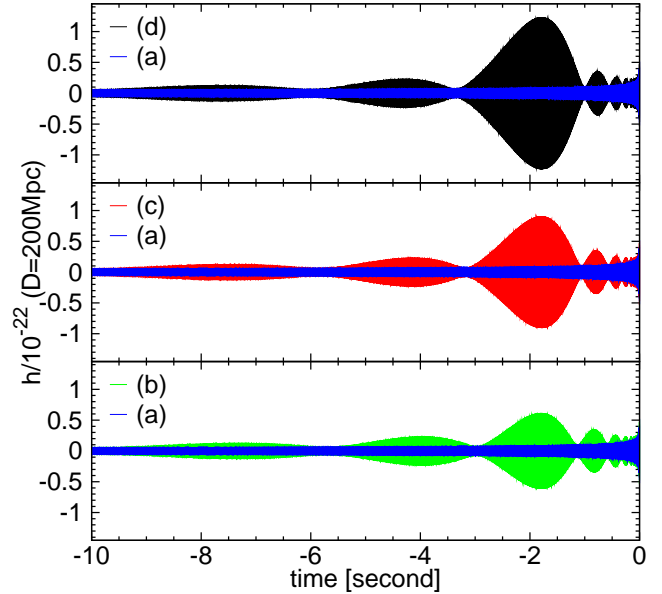}
  \end{center}
  \caption{
The time-domain gravitational waveform $h(t)$.
The parameters are the same as those of Fig.~\ref{fig:hFD_kchange}. 
}%
\label{fig:hTD_kchange}
\end{figure}

The value of $f_{\rm peak}$ and $\tau_{\rm peak}$ for the parameters in Figs. ~\ref{fig:hFD}$-$\ref{fig:hTD} 
is (b) $(67~{\rm Hz}, -6.2~{\rm s})$, 
(c) $(107~{\rm Hz}, -1.8~{\rm s})$, and 
(d) $(169~{\rm Hz}, -0.5~{\rm s})$, respectively.
We can confirm that these values match the location of the highest peaks in Figs. ~\ref{fig:hFD}$-$\ref{fig:hTD} well.

These large deviations of the waveform from GR are produced by the mixing of the two gravitons,
and they depend on the bigravity parameters. 
Thus, these deviations help us put constraints on the bigravity with the GW observations.

The amplitude of the peak is determined by Eq.~(\ref{eq:abshf}). 
The phase difference at the highest peak, which occurs at $x\approx 1$, becomes
\begin{eqnarray}
\Delta \delta\Phi\sim \frac{\sqrt{2} \mu\sqrt{\tilde{c}-1} D_L}
{\sqrt{1+\kappa\xi_c^2}}.
\label{eq:dPhix1}
\end{eqnarray}
For all sets of the bigravity parameters in Fig. \ref{fig:hFD}, $\Delta \delta\Phi$ and $B_1 B_2$ 
at the peak in Eq.~(\ref{eq:abshf}) take the same value. Thus, there is no difference in the amplification of the highest peak
caused by the bigravity effect. 
The difference of these peak amplitudes in Fig.\ref{fig:hFD} is just caused 
by the difference of ${\cal{A}}(f_{\rm peak})$.

In Figs.~\ref{fig:hFD_kchange}$-$\ref{fig:hTD_kchange}, we compare the waveforms with
different values of $\kappa\xi_c^2$  
in the case of $(\mu^2,~\tilde{c}-1)=(10^{-33}~{\rm cm}^{-2},~10^{-18})$.
As can be seen in Eq.~(\ref{eq:fpeak}), 
$f_{\rm peak}$ does not depend on $\kappa \xi_c^2$. 
Thus, the peak frequency does not change at all in Figs.~\ref{fig:hFD_kchange}$-$\ref{fig:hTD_kchange}.
On the other hand, 
we find in Figs.~\ref{fig:hFD_kchange}$-$\ref{fig:hTD_kchange} 
that the deviation of the bigravity waveforms is larger for a larger $\kappa\xi_c^2$.
This can be understood as a consequence of larger value of $\left|B_1B_2\right|$ 
for larger $\kappa\xi_c^2$ in Eq. (\ref{eq:abshf}).

\section{Analysis methods for testing modified gravity theory}\label{sec:statistics}
In this section, we briefly review the methods to test the MG theories. 
Vallisneri~\cite{Vallisneri:2012qq} has proposed a model comparison analysis of simple MG, and derived 
a formula that characterize the possibility to detect the effects of MG on gravitational waves.

First, we define the noise-weighted inner product of signals $h_{\rm A}$ and $h_{\rm B}$ by
\begin{eqnarray}
 (h_{\rm A}|h_{\rm B}) \equiv 4{\rm Re}\int_{f_{\rm min}}^{f_{\rm max}} \frac{h_{\rm A}(f)h_{\rm B}(f)}
 {S_n(f)}df,
\end{eqnarray}
where $S_n(f)$ is the one-sided noise power spectrum density of a detector.

The limits of integration $f_{\rm min}$ and $f_{\rm max}$ are taken to be 
$f_{\rm min}=f_{\rm low}$ and $f_{\rm max}=f_{\rm ISCO}$ 
where $f_{\rm low}$ is the lower cutoff frequency that is defined for each detector, 
while $f_{\rm ISCO}$ is the frequency at the innermost stable circular orbit of the binary.
We adopt $f_{\rm ISCO}=(6^{3/2} \pi M_t)^{-1}$ as an approximation.

The signal-to-noise ratio (SNR) for a given signal $h$ is its norm defined as
\begin{eqnarray}
 {\rm SNR} = |h| = \sqrt{(h|h)}.
\end{eqnarray}
We also define the fitting factor (FF) \cite{Apostolatos:1995pj} 
that is used to characterize the deviation of a MG waveform 
from the GR waveform. The FF is defined as
\begin{eqnarray}
 {\rm FF}(\theta_{\rm MG})=\max_{\theta_{\rm GR}}
 \frac{(h_{\rm GR}(\theta_{\rm GR})|h_{\rm MG}(\theta_{\rm MG}))}
 {|h_{\rm GR}(\theta_{\rm GR})||h_{\rm MG}(\theta_{\rm MG})|}, \label{FF}
\end{eqnarray}
where $h_{\rm GR} (\theta_{\rm GR})$ and $h_{\rm MG}(\theta_{\rm MG})$ are the GR and MG waveforms,
$\theta_{\rm GR}$ represents the source parameters in GR, and  $\theta_{\rm MG}$ represents the parameters 
in the MG theory.

By definition, the maximum of FF is 1, which is realized when the MG waveform coincides with the GR waveform. 
Thus, $1-{\rm FF}$ measures the strength of the MG corrections 
that cannot be absorbed by the variation of the GR source parameters.

The SNR and FF of each waveform in Figs.\ref{fig:hFD}$-$\ref{fig:hTD} 
become as follows: $({\rm SNR}, {\rm FF})=$ (a) $(8.7, 1.0)$, (b) $(31, 0.50)$, (c) $(26, 0.47)$, 
(d) $(21, 0.53)$. 
The same values for Fig. \ref{fig:hFD_kchange} become as follows: 
$({\rm SNR},~{\rm FF})=$ (a) $(8.7,~1.0)$, (b) $(19,~0.58)$, 
(c) $(26,~0.47)$, (d) $(34,~0.41)$.

Now we explain Vallisneri's formula that is based on the Bayesian hypothesis testing.
The Vallisneri's formula can be used for 
estimating the SNR value required for discrimination of gravity models based on FF.
This analysis is valid for large SNR signals and Gaussian detector noise.

In this method, the odds ratio is a key quantity 
that is interpreted as the odds of MG over GR. 
The Bayesian odds ratio for MG over GR is defined as
\begin{eqnarray}
 {\cal O}=\frac{P({\rm MG}|s)}{P({\rm GR}|s)}
 =\frac{P({\rm MG})}{P({\rm GR})} \frac{P(s|{\rm MG})}{P(s|{\rm GR})},
\end{eqnarray}
where $P({\rm MG}|s)$ and $P({\rm GR}|s)$ are the posterior probabilities of the MG and 
GR hypotheses for a given data $s$,
$P({\rm MG})$ and $P({\rm GR})$ are the prior probabilities of the MG and GR 
hypotheses, and $P(s|{\rm MG})$ and $P(s|{\rm GR})$ are the fully marginalized likelihood 
or evidence of the MG and GR hypotheses.
The odds ratio when the data contain a MG signal is given by
${\cal O}_{\rm MG} = P({\rm MG}|s_{\rm MG}) / P({\rm GR}|s_{\rm MG})$,
while the odds ratio when the data contain a {\rm GR} signal is given by
${\cal O}_{\rm GR} = P({\rm MG}|s_{\rm GR}) / P({\rm GR}|s_{\rm GR})$,
where $s_{\rm MG}$ is the data that contain the MG signal
and  $s_{\rm GR}$ is the data that contain the GR signal.
Cornish, {\it et al}.~\cite{Cornish:2011ys} have shown that 
in the limit of large SNR and small MG deviations, 
the logarithm of the odds ratio scales as ${\rm SNR}_{\rm res}^2$,
where the residual signal-to-noise ratio, ${\rm SNR}_{\rm res}$, 
is defined as ${\rm SNR}_{\rm res}\equiv{\rm SNR}\sqrt{1-{\rm FF}}$.
We declare the detection of MG when the odds ratio exceeds a certain threshold 
${\cal O}_{\rm thr}$. We set the threshold ${\cal O}_{\rm thr}$ 
by requiring a given false alarm probability, $F$, 
which is the fraction of observation
in which ${\cal O}$ happens to exceed ${\cal O}_{\rm thr}$ in the case of GR signal.
The efficiency of the detection, $E$, is the fraction of observation 
in which ${\cal O}$ exceeds ${\cal O}_{\rm thr}$
in the case of MG signal. 
When one computes $E$ as a function of $F$, $E$ is a simple function of the residual 
signal-to-noise ratio ${\rm SNR}_{\rm res}$.

The formula is given as \cite{Vallisneri:2012qq}
\begin{eqnarray}
 E=1-\frac{1}{2}( {\rm erf}(-{\rm SNR}_{\rm res}+{\rm erfc}^{-1}(F)) \nonumber\\
 -{\rm erf}(-{\rm SNR}_{\rm res}-{\rm erfc}^{-1}(F)) ), \label{eq:Effi}
\end{eqnarray}
where
$z={\rm erfc}^{-1}(F)$ is the solution of ${\rm erfc}(z)=F$. 
In this paper, we assume $E=1/2$ and $F=10^{-4}$. 
The solution of (\ref{eq:Effi}), with $E=1/2$ and $F=10^{-4}$, is denoted as 
${\rm SNR}_{\rm res}={\rm SNR}_{\rm res}^{c}$.
The SNR required for confident MG detection is then given as 
${\rm SNR}_{\rm req}={\rm SNR}_{\rm res}^{c}/\sqrt{1-{\rm FF}}$.
We can find that the {\rm SNR} required to detect $10\%$ of deviations from GR (${\rm FF}=0.9$) is $8.699$.

Del Pozzo {\it et al}. \cite{DelPozzo:2014cla} have shown that 
the scaling that the logarithm of the odds ratio scales as ${\rm SNR}_{\rm res}^2$ 
holds in the case of two or more MG parameters 
at the lowest order of $(1-{\rm FF})^2$.
Thus, Eq. (\ref{eq:Effi}) holds for two or more MG parameters.

When the bigravity signal is detected, the next question is how accurately the bigravity parameters 
can be measured. To quantify the measurement accuracy of parameters, we compute the standard Fisher matrix,
\begin{eqnarray}
 \Gamma_{ab} \equiv \left( \left. \frac{\partial h}{\partial \theta^a} \right| \frac{\partial h}{\partial \theta^b} \right), 
\end{eqnarray}
which is an $8\times8$ matrix in the present context.
For sufficiently strong signal, 
the measurement accuracy of a parameter $\theta^a$ can be evaluated as
\begin{eqnarray}
 \Delta\theta^a \equiv \sqrt{\left<(\theta^a-\left<\theta^a\right>)^2\right>}=\sqrt{(\Gamma^{-1})^{aa}}\,.
\end{eqnarray}

\section{Phenomenological model}
\label{sec:phenom}
First, we consider the phenomenological model, in which the bigravity parameters 
$\mu^2$, $\tilde{c}-1$, and $\kappa\xi_c^2$ are treated as independent parameters,
although $\mu^2$ and $\tilde{c}-1$ are related with each other in the case of the ghost-free
bigravity. This case is discussed in the succeeding section.

\subsection{Detectability of the bigravity corrections to the waveforms}
\label{sec:detectability}
In this section, we evaluate the detectable region of the parameters of the bigravity theory 
with the observation of gravitational waves by an advanced laser interferometer. 
We consider the three cases of binary with masses, $(1.4M_\odot,1.4M_\odot)$ ($f_{\rm ISCO}=1570$Hz), 
$(1.4M_\odot,10M_\odot)$ ($f_{\rm ISCO}=386$Hz), and 
$(10M_\odot,10M_\odot)$ ($f_{\rm ISCO}=219$Hz).
In this paper, we consider the face-on binaries that are located at the zenith direction from the detector. 
We thus do not consider the dependence on the inclination, the source location on the sky, 
and the polarization angle of the wave.

\begin{figure}[tbp]
  \begin{center}
   \includegraphics[width=0.50\textwidth,angle=0]{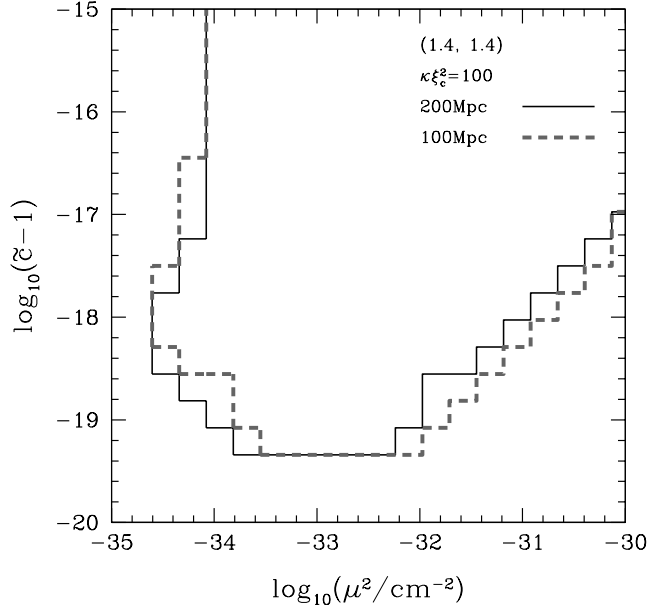}
  \end{center}
  \caption{
The detectable region of the bigravity corrections to the waveforms
in the case $(m_1,m_2)=(1.4\solM, 1.4\solM)$ and $\kappa\xi_c^2=100$.
Curves correspond to the distance to the source at $D_L=200~{\rm Mpc}$ (solid) and $100~{\rm Mpc}$ (dashed).
The detectable region is upper and right-hand side of these curves. 
The detectable region is defined as the region where 
${\rm SNR}>{\rm SNR}_{\rm req}$ is satisfied. 
The false-alarm probability is set to $F=10^{-4}$
}%
\label{fig:Detectable1414_200_100Mpc}
\end{figure}

We obtain ${\rm SNR}_{\rm res}$ from Eq.~(\ref{eq:Effi}) by setting $E=1/2$ and $F=10^{-4}$.
The detectable region of the bigravity correction is the region 
where ${\rm SNR}>{\rm SNR}_{\rm req}={\rm SNR}_{\rm res}/\sqrt{1-{\rm FF}}$ is satisfied. 
Figure~\ref{fig:Detectable1414_200_100Mpc} shows the detectable region 
of $(\mu^2,~\tilde{c}-1)$ in the case of
$(m_1,m_2)=(1.4M_\odot,1.4M_\odot)$ and $\kappa \xi_c^2=100$.
Curves correspond to the distance to the source $D_L=200~{\rm Mpc}$ (solid line) and $100~{\rm Mpc}$ 
(dashed line), respectively. The upper-right regions of these lines are the region in which 
the bigravity correction is detectable. 
The regions shown in Fig.~\ref{fig:Detectable1414_200_100Mpc}
have not been excluded 
with the solar system experiments yet (see Ref.~\cite{DeFelice:2013nba} for detail).
Thus, this figure shows an interesting possibility to constrain and detect the bigravity correction 
to the GR waveforms from CCB.

By comparing the regions in Fig.~\ref{fig:Detectable1414_200_100Mpc}, 
we find that the detectable region for $D_L=100~{\rm Mpc}$ is slightly larger 
than that for $D_L=200~{\rm Mpc}$. The effect of larger SNR for smaller distance 
turns out not to be very large.

\begin{figure}[tbp]
  \begin{center}
   \includegraphics[width=0.50\textwidth,angle=0]{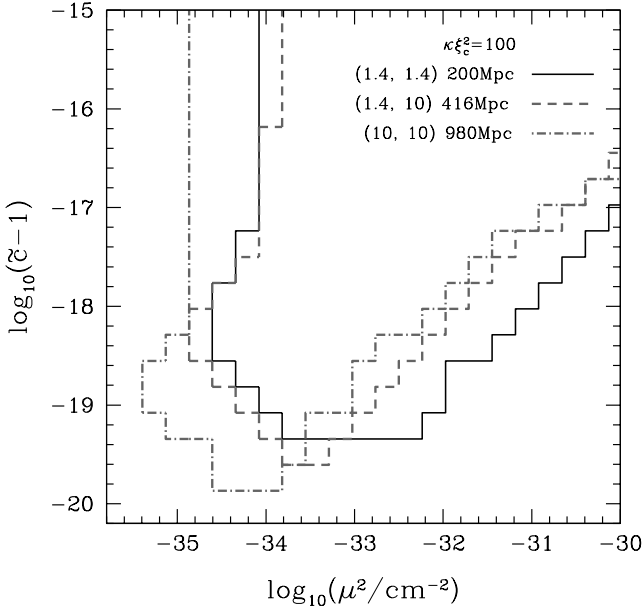}
  \end{center}
  \caption{
A plot similar to Fig.~\ref{fig:Detectable1414_200_100Mpc}  but for the waveforms
from BNS with $(m_1,m_2)=(1.4\solM, 1.4\solM)$ at 200Mpc (solid),
NSBH with $(m_1,m_2)=(1.4\solM, 10\solM)$ at 416Mpc (dashed),
and BBH with $(m_1,m_2)=(10\solM, 10\solM)$ at 980Mpc (dot-dashed),
respectively. We set $\kappa\xi_c^2=100$.
The detectable region is upper and right-hand side of these curves. 
SNR of the gravitational waves from these systems in GR limit are 8.7.
}%
\label{fig:Detectable1414_1410_1010_Rescale}
\end{figure}

\begin{figure}[tbp]
  \begin{center}
   \includegraphics[width=0.50\textwidth,angle=0]{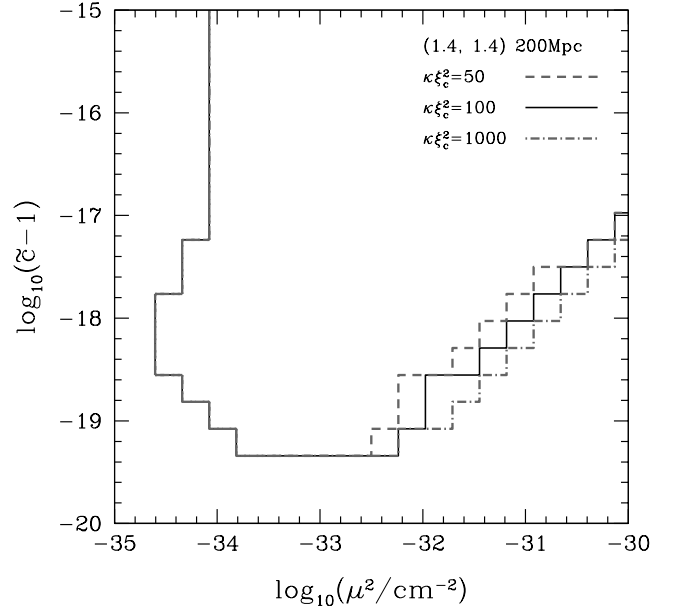}
  \end{center}
  \caption{
A plot similar to Fig.~\ref{fig:Detectable1414_200_100Mpc}, 
but for $\kappa\xi_c^2=50$ (dashed), 100 (solid), and $1000$ (dot-dashed), respectively. 
The masses are $(m_1,m_2)=(1.4\solM, 1.4\solM)$ and the distance is 200Mpc.
}%
\label{fig:Detectable_k50_100_1000}
\end{figure}

We compare the effect of the masses of the binaries on the detectable region. 
We consider NSBH with $(m_1,~m_2)=(1.4\solM,~10\solM)$ and BBH with 
$(m_1,~m_2)=(10\solM,~10\solM)$.
We set the distance of these systems so that the SNR in the GR limit is 8.7,
which is the value for BNS at 200Mpc. 
The distance with ${\rm SNR}=8.7$ becomes 416Mpc for NSBH and 980Mpc for BBH. 
The upper and right-hand side of the lines in
Fig.~\ref{fig:Detectable1414_1410_1010_Rescale}  represents the detectable regions 
on $(\mu^2,~\tilde{c}-1)$ plane. 
For simplicity, we do not consider the cosmological redshift effect. 
We find that, the detectable region in the case of NSBH is slightly smaller than that of BNS. 
On the other hand, the detectable region is slightly larger for BBH than for BNS.

We also consider the cases with different values of $\kappa\xi_c^2$. 
In Fig.~\ref{fig:Detectable_k50_100_1000},
we show the detectable region for $\kappa\xi_c^2=50, 100$, and $1000$ 
for BNS at 200Mpc. 
We find that the detectable region does not strongly depend on the parameter $\kappa\xi_c^2$.

\subsection{Interpretation of the detectable region}
Now, we investigate the origin of the shape of the detectable region 
in Figs.~\ref{fig:Detectable1414_200_100Mpc}$-$\ref{fig:Detectable_k50_100_1000}. 
Eq. (\ref{eq:fpeak}) represents the peak frequency of amplitude of the bigravity waveform 
in the frequency domain 
as a function of $\tilde{c}-1$ and $\mu^2$. 
We recover the dimension and rewrite Eq. (\ref{eq:fpeak}) as
\begin{eqnarray}
\frac{\tilde{c}-1}{\mu^2}=1.1\times 10^{15} \left( \frac{100~{\rm Hz}}{f_{\rm peak}} \right)^2 {\rm cm}^2. 
\label{mc1}
\end{eqnarray}
When the value of $f_{\rm peak}$ is located within the detector's sensitivity band,
and less than $f_{\rm ISCO}$, the bigravity effects can be detected easily. 
We take the maximum frequency of the detector's sensitivity band 
to be $1000$Hz corresponding to the sensitivity curve of advanced LIGO used in this paper. 
Then, the above equation becomes
\begin{eqnarray}
\tilde{c}-1\gtrsim 1.1\times 10^{-19} \left(\frac{\mu^2}{10^{-32}{\rm cm}^{-2}}\right) 
\left( \frac{10^3~{\rm Hz}}{f_{\rm max}} \right)^2. \label{mc1b}
\end{eqnarray}
We can see that this equation approximately expresses the lower boundary of the region 
for $\mu^2>10^{-32}$ cm$^{-2}$  in Fig.~\ref{fig:Detectable1414_200_100Mpc}.

As discussed in Sec. \ref{sec:modwave}, the largest effect of bigravity model 
can occur when $x\approx1$.
In such a case, Eq.~(\ref{deltaPhi}) is rewritten as 
\begin{eqnarray}
\tilde{c}-1&\simeq&1.3\times10^{-18} ~(\Delta\delta\Phi)^2 
\left(\frac{10^{-34} {\rm cm}^{-2}}{\mu^2} \right)
\left( \frac{\kappa\xi_c^2}{100} \right) \nonumber \\
& & \times \left( \frac{200~{\rm Mpc}}{D_L} \right)^2. \label{mc2} 
\end{eqnarray}
If $\Delta\delta\Phi\neq 0$, the deviation of bigravity from GR becomes possible to detect. 
By setting $(\Delta\delta\Phi)\sim 0.3$, 
we can see that Eq. (\ref{mc2}) roughly represents the lower boundary of the detectable region 
for $\mu^2\lesssim 10^{-34} {\rm cm}^{-2}$ in Fig.~\ref{fig:Detectable1414_200_100Mpc}.

We can also eliminate $\tilde{c}-1$ or $\mu^2$ from Eqs. (\ref{mc1}) and (\ref{mc2}).
We obtain
\begin{eqnarray}
  \mu^2 &\simeq&3.4\times10^{-35} ~{(\Delta\delta\Phi)} \left(\frac{f}{10~{\rm Hz}}\right) 
  \left(\frac{\kappa\xi_c^2}{100}\right)^{1/2} \nonumber \\
  & & \times \left(\frac{200~{\rm Mpc}}{D_L}\right)~{\rm cm}^{-2}, \label{eq:mu2} \\
  \tilde{c}-1 &\simeq& 3.9\times10^{-20} (\Delta\delta\Phi) \left(\frac{1000~{\rm Hz}}{f}\right) 
  \left(\frac{\kappa\xi_c^2}{100}\right)^{1/2} \nonumber \\
  & & \times \left(\frac{200~{\rm Mpc}}{D_L}\right). \label{eq:tildec}
\end{eqnarray}
These two equations can give the lower boundary for $\mu^2$ and $\tilde{c}-1$. 
By setting $f\sim f_{\rm min}\sim {\rm a~few} ~10$~Hz for (\ref{eq:mu2}),
$f\sim f_{\rm max}\sim  10^3$~Hz for (\ref{eq:tildec}), 
and $(\Delta\delta\Phi)\sim 0.3$, 
we can see that these two equations represent approximately the lower bound 
of the detectable region for $\mu^2$ and $\tilde{c}-1$
in Fig.~\ref{fig:Detectable1414_200_100Mpc}.

The boundary of Fig.~\ref{fig:Detectable1414_1410_1010_Rescale} 
can be understood similarly. 
The lower boundary of Fig.~\ref{fig:Detectable1414_1410_1010_Rescale} 
is determined by Eq.~(\ref{mc1b}). 
For these systems we have $f_{\rm ISCO}=1570$Hz (BNS), 
386Hz (NSBH), and 220Hz (BBH). 
Since $f_{\rm ISCO}$ for NSBH and BBH becomes lower than that for BNS, 
$f_{\rm max}$ in Eq.~(\ref{mc1b}) becomes smaller, which raises the lower boundary 
for $\mu^2$ to $\mu^2>10^{-33}$ cm$^{-2}$ in Fig.~\ref{fig:Detectable1414_1410_1010_Rescale}.

Other differences are produced by the difference of distance in Eq.~(\ref{mc2}). 
For NSBH and BBH, the distance is larger and the lower boundary becomes lower than that of BNS.
We can also understand most of the lowest boundary of $\mu^2$ and $\tilde{c}-1$ 
in Fig.~\ref{fig:Detectable1414_1410_1010_Rescale} 
from the dependence on the distance of Eqs.~(\ref{eq:mu2})$-$(\ref{eq:tildec}).
However, the difference between BNS and NSBH of the lowest boundary for $\mu^2$ 
is very small. 

In Fig.~\ref{fig:Detectable_k50_100_1000}, we see that the difference of $\kappa\xi_c^2$ produces 
only a small difference in the detectable region. 
As we saw in Figs.~\ref{fig:hFD_kchange}$-$\ref{fig:hTD_kchange}, 
the amplitude of bigravity waveform becomes larger when $\kappa\xi_c^2$ is larger. 
Thus, SNR of the signal becomes larger. 
However, from  Eqs.~(\ref{mc2})$-$(\ref{eq:tildec}), 
we find that larger $\kappa\xi_c^2$ raises the lower boundary of $\mu^2$ and $\tilde{c}-1$.
These two  effects compensate each other, and the difference of the detectable region 
becomes very small in Fig.~\ref{fig:Detectable_k50_100_1000}.
The only difference we can see is the boundary for $\mu^2>10^{-32}$cm$^{-2}$, 
for which Eq.~(\ref{mc1b}) determines the boundary. 
Since Eq.~(\ref{mc1b}) dose not depend on $\kappa\xi_c^2$, large SNR for larger $\kappa\xi_c^2$
produces slightly wider detectable region.

\begin{figure}[htbp]
  \begin{center}
    \includegraphics[keepaspectratio=true,height=80mm]{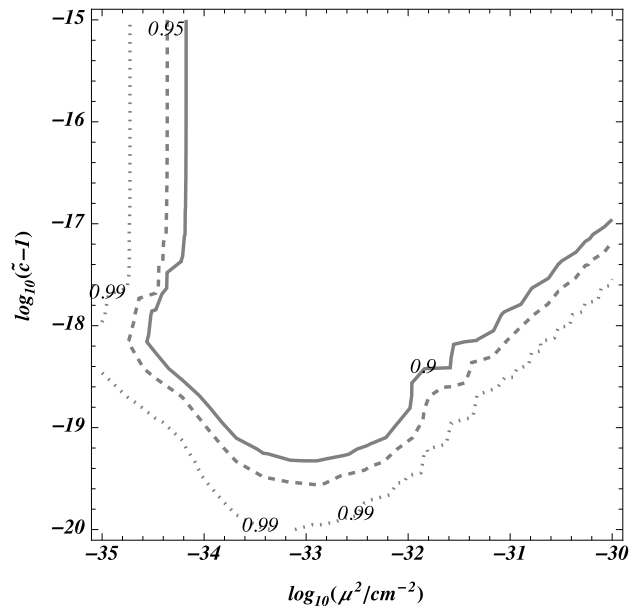}
  \end{center}
  \caption{
Contour plots of the fitting factor between the GR and bigravity waveforms
in the $(\mu^2,~\tilde{c}-1)$ parameter space. 
Here we adopt the model $\kappa\xi_c^2=100$.
Curves correspond to contours of ${\rm FF}=0.9$ (solid), ${\rm FF}=0.95$ (dashed), 
and ${\rm FF}=0.99$ (dotted).
We assume BNS at $D_L=200~{\rm Mpc}$.
}%
\label{fig:FF_1414}
\end{figure}

Here, we mention the correspondence between Fig.~\ref{fig:Detectable1414_200_100Mpc}
and the contours of the fitting factor between the GR and bigravity waveforms, 
which are plotted in Fig.~\ref{fig:FF_1414}. 
The FF is computed by maximizing Eq.~(\ref{FF}) with respect to 
$m_1$ and $m_2$ for each value of ($\mu^2$,~$\tilde{c}-1$), at fixed $\kappa\xi_c^2=100$.
We find that the detectable region of the bigravity corrections in Fig.~\ref{fig:Detectable1414_200_100Mpc} 
is very similar to the red solid contour of ${\rm FF}=0.9$ in Fig.~\ref{fig:FF_1414}. 
This fact shows that the detectable region in Fig.~\ref{fig:Detectable1414_200_100Mpc} is almost determined 
by the value of the fitting factor in this case. 

\begin{figure}[htbp]
  \begin{center}
    \includegraphics[keepaspectratio=true,height=80mm]{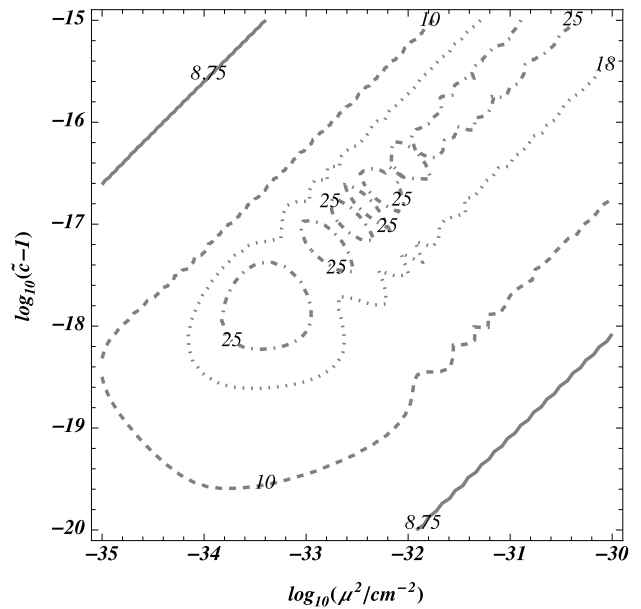}
  \end{center}
  \caption{
Contour plots of the SNR of bigravity waveforms
in the $(\mu^2,~\tilde{c}-1)$ parameter space. 
The parameters are the same as those of Fig.~\ref{fig:Detectable1414_200_100Mpc}.
Curves correspond to contours of ${\rm SNR}=8.75$ (solid),  
${\rm SNR}=10$ (dashed), ${\rm SNR}=18$ (dotted), and ${\rm SNR}=25$ (dot-dashed).
We assume BNS at $D_L=200~{\rm Mpc}$.
}%
\label{fig:SNRcont_1414}
\end{figure}

Figure~\ref{fig:SNRcont_1414} shows the contour of SNR for BNS. 
By comparing ${\rm SNR}_{\rm req}$ from Fig.~\ref{fig:FF_1414} and SNR from Fig.~\ref{fig:SNRcont_1414}, 
we can obtain the detectable region of Fig.~\ref{fig:Detectable1414_200_100Mpc}
as the region where ${\rm SNR}>{\rm SNR}_{\rm req}$ is satisfied.

\subsection{Constraints on bigravity parameters }

\begin{figure}[htbp]
  \begin{center}
   \includegraphics[width=0.50\textwidth,angle=0]{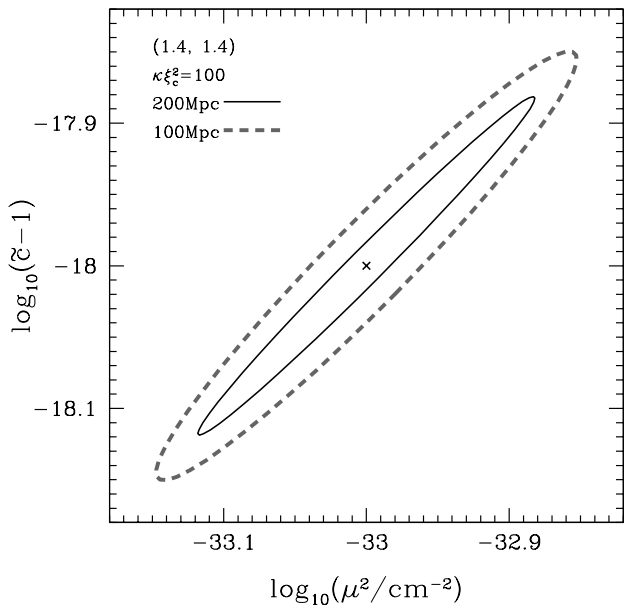}
  \end{center}
  \caption{
Projected $1\sigma$ error contours on the $(\mu^2$,~$\tilde{c}-1)$ plane.
The results are obtained from the Fisher matrix with 8-parameters, 
$\log\mu^2, ~\log(\tilde{c}-1), ~\kappa\xi_c^2, ~\log D_L, ~{\cal M}, ~\eta, ~t_c$, and $\Phi_c$,
and marginalized over 6 parameters other than $\log\mu^2$ and $\log(\tilde{c}-1)$.
The fiducial model is
$(\mu^2, \tilde{c}-1)=(10^{-33} {\rm cm}^{-2}, 10^{-18})$,
for BNS at $D_L=200~{\rm Mpc}$ (solid) and at $100~{\rm Mpc}$ (dashed).
SNR is renormalized to ${\rm SNR}=10$.
}%
\label{fig:cont_m33_c18_SNR10}
\end{figure}

\begin{figure}[htbp]
  \begin{center}
   \includegraphics[width=0.50\textwidth,angle=0]{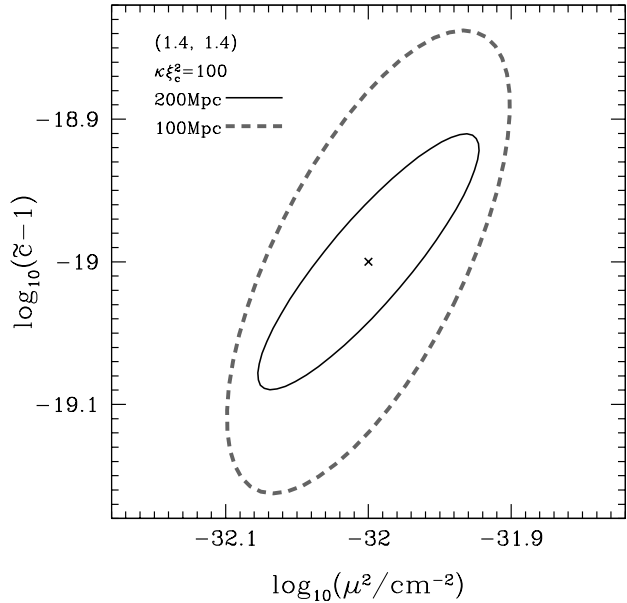}
  \end{center}
  \caption{
Same as Fig.~\ref{fig:cont_m33_c18_SNR10} but for
the fiducial model, $(\mu^2, \tilde{c}-1)=(10^{-32} {\rm cm}^{-2}, 10^{-19})$.
SNR is renormalized to ${\rm SNR}=10$.
}%
\label{fig:cont_m32_c19_SNR10}
\end{figure}

Next, we evaluate the measurement accuracy of the bigravity parameters.
We compare the error contour on the $(\mu^2,~\tilde{c}-1)$ plane 
for the sources at different distances. 
In order to see the genuine effect of the bigravity on the waveform through the different source distance,
we renormalize the amplitude of the waveforms so that the signals have the same SNR. 
In Fig.~\ref{fig:cont_m33_c18_SNR10}, we show the measurement accuracy 
in the case of $(\mu^2,~\tilde{c}-1)=(10^{-33}~{\rm cm}^{-2},~10^{-18})$, 
and for the BNS at $200$ and $100~{\rm Mpc}$, but with SNR renormalized to ${\rm SNR}=10$. 
In this case, the expected accuracy of $\log\mu^2$ is ${\cal O}(0.1)\%$ at $1\sigma$ level. 
We find that the accuracy is better for the $200~{\rm Mpc}$ case. 
Note that the phase shift, $\delta\Phi_{1,2}$, in Eq. (\ref{deltaPhi}) depends on the distance.
For the parameters in Fig.~\ref{fig:cont_m33_c18_SNR10}, 
The factor $1+2B_1B_2(\cos(\Delta\delta\Phi)-1)$ is $97.1$ for $D_L=200~{\rm Mpc}$ and
$41.1$ for $D_L=100~{\rm Mpc}$. 
Thus, the bigravity effect is larger for the $200~{\rm Mpc}$ case. 
In Fig. \ref{fig:cont_m32_c19_SNR10}, we show the error contour in the case of 
different parameters of $(\mu^2,~\tilde{c}-1)=(10^{-32}~{\rm cm}^{-2},~10^{-19})$.
We find the same trend as above:
the $1\sigma$ error of $\log\mu^2$ is ${\cal O}(0.1)\%$, 
and the accuracy is better for the $200~{\rm Mpc}$ case.

\section{The DFNT subset of the bigravity model}
\label{sec:DFNT}
\begin{figure}[thbp]
  \begin{center}
   \includegraphics[width=0.50\textwidth,angle=0]{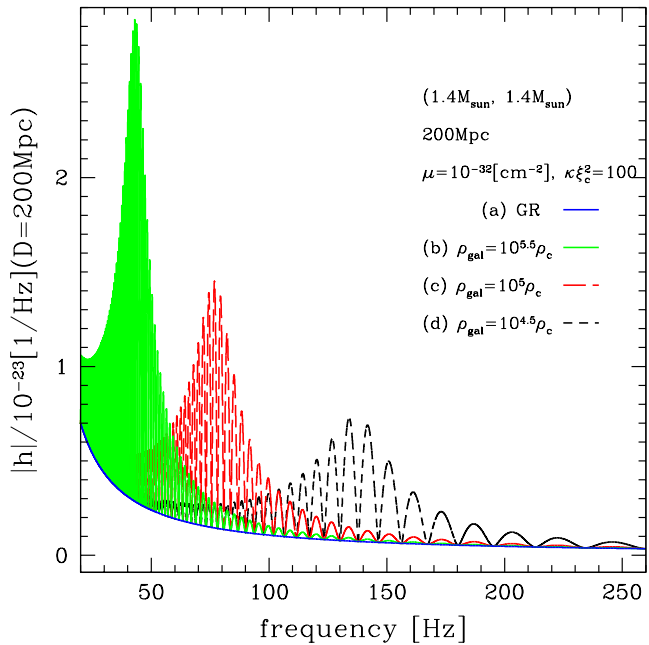}
  \end{center}
  \caption{
The frequency-domain gravitational waves $h(f)$ for DFNT subset of the bigravity model
for different values of the average density in the galaxies $\rho_{\rm gal}$, where GWs are generated.
The curves are plotted for (a) GR (solid (blue)) and for the DFNT subset of the bigravity model with 
(b) $\rho_{\rm gal}=10^{5.5}\rho_{\rm c}$ (dot-dashed (green)),
(c) $10^{5}\rho_{\rm c}$ (long-dashed (red)), and (d) $10^{4.5}\rho_{\rm c}$ 
(dashed (black)), respectively, at fixed $(\mu^2,~\kappa\xi_c^2)=(10^{-32}~{\rm cm}^{-2},~100)$.
Here we consider BNS at the distance, $D_L=200~{\rm Mpc}$. 
The SNR and the fitting factor between GR waveform and  each waveform in this figure 
become as follows. $({\rm SNR}, {\rm FF})=$ (a) $(8.7, 1.0)$, (b) $(26, 0.71)$, (c) $(24, 0.72)$, (d) $(19, 0.73)$.
}%
\label{fig:hFD_DFNT}
\end{figure}

Next, we study the DFNT subset of the bigravity model~\cite{DeFelice:2013nba},  
in which the bigravity parameters obey the relation
\begin{eqnarray}
 \tilde{c}-1=3H_0^2 \frac{\rho_{\rm m}}{\rho_{c}} \frac{1+\kappa\xi_c^2}{\mu^2},
\label{eq:ctildeDFNT}
\end{eqnarray}
where $H_0$ is the Hubble parameter at the present epoch and $\rho_{\rm c}$ is the critical density. 
The value of $\tilde{c}-1$ is large in the high density region, 
while it is small in the low density region. 
We assume GWs are generated in a galaxy where the density is higher than the average density 
in the intergalactic space. 
We also assume that GWs experience much lower density during the propagation between galaxies. 
We neglect the effect of the high density region on the phase corrections
$\delta\Phi_{1,2}$, and we evaluate the phase corrections by using the background density of the Universe. 
On the other hand, we assume that
the dispersion relations of the modes 1, 2 adiabatically evolve because of the slow evolution of the background.
Therefore, by assuming conservation of energy for each mode, we
evaluate the amplitude corrections $B_{1,2}$ with the average density in the galaxy, 
$\rho_{\rm gal}$, where binaries are embedded.  
Figure~\ref{fig:hFD_DFNT} shows the gravitational waveforms for the DFNT subset of the bigravity model for 
different values of the average density in the galaxy.
Curves in Fig.~\ref{fig:hFD_DFNT} are for (a) GR [solid (blue)] and for the DFNT subset of the bigravity model with
$\rho_{\rm gal}=$ (b) $10^{5.5}\rho_{\rm c}$ [dot-dashed (green)],
(c) $10^{5}\rho_{\rm c}$ [long-dashed (red)], and (d) $10^{4.5}\rho_{\rm c}$ [dashed (black)], respectively.
We set $(\mu^2,~\kappa\xi_c^2)=(10^{-32}~{\rm cm^{-2}},~100)$ and $D_L=200~{\rm Mpc}$.
The gravitational waveforms for the DFNT subset of the bigravity model
are significantly different from those for the phenomenological bigravity model.
From Eqs.~(\ref{eq:fpeak}) and (\ref{eq:ctildeDFNT}),
we see that $f_{\rm peak}$ increases as $\mu^2$ increases, $\kappa\xi_c^2$ decreases, 
and $\rho_{\rm gal}$ decreases, and does not depend on $D_L$.
The value of $f_{\rm peak}$ for the parameters in Fig.~\ref{fig:hFD_DFNT} are (b) $44~{\rm Hz}$, 
(c) $78~{\rm Hz}$, and (d) $138~{\rm Hz}$.
The SNR and FF of each waveform in Fig.\ref{fig:hFD_DFNT} 
become as follows: $({\rm SNR}, {\rm FF})=$ (a) $(8.7, 1.0)$, (b) $(26, 0.71)$, (c) $(24, 0.72)$, 
(d) $(19, 0.73)$. 

\begin{figure}[tbp]
  \begin{center}
   \includegraphics[width=0.50\textwidth,angle=0]{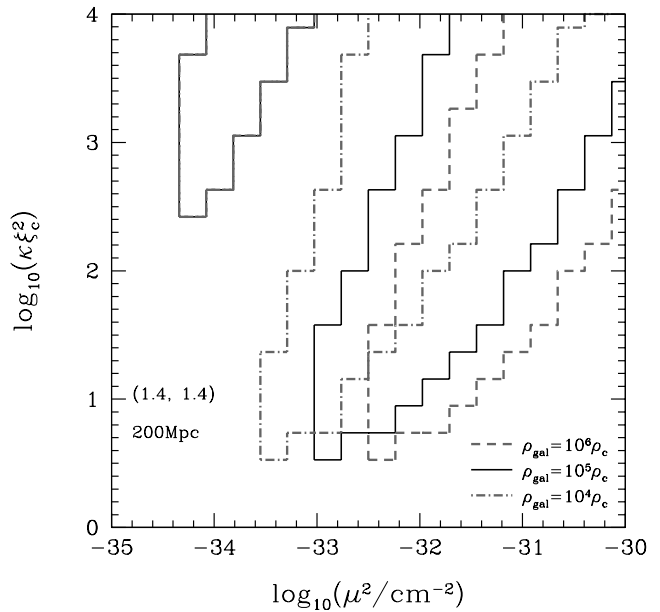}
      \end{center}
  \caption{
The detectable region of the bigravity corrections to the waveforms for DFNT subset of the bigravity model
in the case $(m_1,m_2)=(1.4\solM, 1.4\solM)$ and $D_L=200~{\rm Mpc}$.
Curves correspond to the average density in the galaxies 
$\rho_{\rm gal}=10^{5.5}\rho_{\rm c}$ (dashed), $10^{5}\rho_{\rm c}$ (solid), and 
$10^{4.5}\rho_{\rm c}$ (dot-dashed).
}%
\label{fig:Detectable_DFNT_density}
\end{figure}

Figure~\ref{fig:Detectable_DFNT_density} shows the detectable region of 
$(\mu^2,~\kappa\xi_c^2)$ for the DFNT subset of the bigravity model in the case of $(m_1,m_2)=(1.4\solM, 1.4\solM)$ and
$D_L=200~{\rm Mpc}$.
Curves correspond to the average density in the galaxies 
$\rho_{\rm gal}=10^{5.5}\rho_{\rm c}$ (dashed),
$\rho_{\rm gal}=10^{5}\rho_{\rm c}$ (solid), and
$\rho_{\rm gal}=10^{4.5}\rho_{\rm c}$ (dot-dashed), respectively.
There are two detectable regions.
The right region corresponds to the region where the amplitude deviation from that of the GR waveform is significant, 
while the left region corresponds to the region where the phase deviation from that of the GR waveform is significant.
The left region does not exist in the phenomenological model. 
As an example, if we pick up one point in the left region 
at $(\mu^2,~\kappa\xi_c^2)=(10^{-34}~{\rm cm^{-2}},~10^{3.2})$, 
we have $f_{\rm peak}=0.20$ Hz for $\rho_{\rm Gal}=10^5\rho_{\rm c}$, 
which is out of the detector sensitivity band. 
While the amplitude and ${\rm SNR} (=8.7)$ is very similar to that in GR waveform in this case,
the phase corrections help us detect the bigravity corrections. 
In this case, ${\rm FF}=0.63$. 
The left region does not depend on the average density of the galaxy 
because the phase corrections $\delta\Phi_{1,2}$ do not depend on $\rho_{\rm gal}$.
Thus, all three lines overlap each other. 

\begin{figure}[tbp]
  \begin{center}
   \includegraphics[width=0.50\textwidth,angle=0]{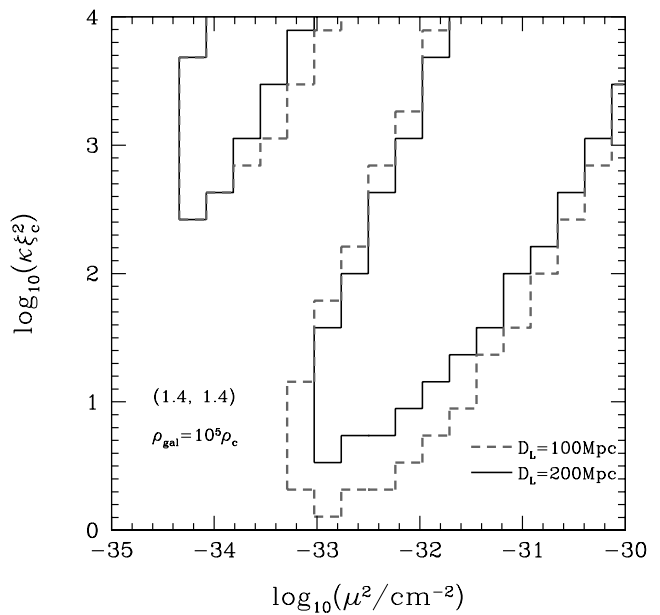}
  \end{center}
  \caption{
A plot similar to Fig.~\ref{fig:Detectable_DFNT_density},
but for $D_L=100~{\rm Mpc}$ (dashed) and $200~{\rm Mpc}$ (solid), respectively.
The masses are $(m_1,m_2)=(1.4\solM, 1.4\solM)$.
We set $\rho_{\rm gal}=10^{5}\rho_{\rm c}$.
}%
\label{fig:Detectable_DFNT_dist}
\end{figure}

We also consider the effect of the distance to the source on the detectable region.
Figure~\ref{fig:Detectable_DFNT_dist} shows the detectable region for $D_L=100~{\rm Mpc}$ 
and $200~{\rm Mpc}$ for BNS. 
The detectable region for $D_L=100~{\rm Mpc}$ is 
slightly larger than that for $D_L=200~{\rm Mpc}$.
This is because of larger SNR for smaller distance.

\section{Summary and conclusions}
\label{sec:summary}
In this paper, we investigated the detectability
of the ghost-free bigravity theory 
with the observation of gravitational waves from inspiraling compact binaries.
Graviton oscillations generate deviations of the gravitational waveform from that of  GR.
These effects can be used to
put constraints on the bigravity model. 

We calculated modified inspiral waveforms and 
observed the amplitude modulation due to graviton oscillations
in the phenomenological model and in the DFNT subset of the bigravity model. 
We found that there is a characteristic frequency for the peak of the amplitude of the inspiral waveforms
that is determined by the bigravity parameters.

In order to assess the detectability of the deviation of the waveform from GR prediction 
due to bigravity effects, we used the formula derived by Vallisneri 
that is based on the  Bayesian hypothesis testing, 
and which uses the fitting factor to compute the Bayesian odds ratio. 
With this method, 
we evaluated the detectability of the deviations of the waveforms by an advanced laser interferometer.
We found that there is a region of the parameter space  of the bigravity model where 
the deviation can be detected. 
The detectable region corresponds to 
the effective graviton mass of $\mu^2 \simgt 10^{-34}~{\rm cm}^{-2}$, 
and the propagation speed of the hidden graviton mode of $\tilde{c}-1 \simgt 10^{-19}$
for the phenomenological model, and $\mu^2 \simgt 10^{-34}~{\rm cm}^{-2}$ and 
$\kappa\xi_c^2 \simgt 10^{0.5}$ for the DFNT subset of the bigravity model.

The shape of the detectable region can be easily understood by using the formula that describes 
the bigravity correction to the waveform.
The existence of the detectable region is rather robust and
is not strongly affected by the source parameters within the region of interest.
We thus conclude that 
GW observations can be a powerful probe of graviton oscillations.

In the phenomenological model, 
we also studied the possibility to constrain the bigravity parameters that characterize 
graviton oscillations 
by the observations of the GW from binary inspirals.
We found that accuracy in determining the effective graviton mass $\log\mu^2$ is
${\cal O}(0.1)\%$ for the particular model with $(\mu^2,~\tilde{c}-1)=(10^{-33}~{\rm cm}^{-2},~10^{-18})$.
We also investigated the dependence of the accuracy on binary masses and the distance to the source.

In this paper, we fixed the distance to the source when we calculated the FF.
In the real data analysis,  
it is possible to determine the distance as well as the direction to the source and the inclination angle
by using a network of GW detectors. 
Even in that case, it would be very helpful 
if electromagnetic follow-up observations could determine the distance by identifying the host galaxy.
Also, we have not included the spins of the stars in the binaries. 
If the spin precession effect exists, there will be an amplitude modulation due to the spin precession effect. 
Such modulation will be mixed with the modification caused by the bigravity effects, and 
the waveform will become more complicated. 
In such a case, the results in this paper may be changed. 
Since the spin may not be neglected for black holes, it is important to investigate the effects of spin.
We plan to investigate it in the future.

If we consider future detectors such as 
Einstein Telescope~\cite{Punturo:2010}, 
eLISA/NGO~\cite{AmaroSeoane:2012je}, or DECIGO/BBO
~\cite{Seto:2001qf,Kawamura:2006up,Kawamura:2011zz},
it will be possible to constrain another region 
because it will be possible to detect GWs from coalescing binaries at much larger distance, 
and at a different frequency region.
We also plan to investigate such cases in the future.

\acknowledgments
This work was supported by MEXT Grant-in-Aid for Scientific Research on Innovative Areas, 
"New Developments in Astrophysics Through Multi-Messenger Observations of Gravitational Wave Sources," 
Grants No. 24103005 and No. 24103006.
This work was also supported in part by Grant-in-Aid for Scientific Research
(C) Grants No. 23540309, (A) No. 24244028, and (B) No. 26287044. 
This work was also supported by JSPS Core-to-Core Program, A. Advanced Research Networks. 




\begin{thebibliography}{99}
\bibitem{Harry:2010zz} 
  G.~M.~Harry (LIGO Scientific Collaboration),
  Advanced LIGO: The next generation of gravitational wave detectors,
  Classical Quantum Gravity {\bf 27}, 084006 (2010).
  
\bibitem{Virgo}
 https://tds.ego-gw.it/ql/?c=6589.

\bibitem{Somiya:2011np} 
  K.~Somiya (KAGRA Collaboration),
  Detector configuration of KAGRA: The Japanese cryogenic gravitational-wave detector,
  Classical Quantum Gravity {\bf 29}, 124007 (2012)
  [arXiv:1111.7185 [gr-qc]].
  
\bibitem{Aso:2013eba} 
  Y.~Aso, Y.~Michimura, K.~Somiya, M.~Ando, O.~Miyakawa, T.~Sekiguchi, D.~Tatsumi, and H.~Yamamoto (KAGRA Collaboration),
  Interferometer design of the KAGRA gravitational wave detector,
  Phys.\ Rev.\ D {\bf 88}, 043007 (2013)
  [arXiv:1306.6747 [gr-qc]].

\bibitem{Yunes:2013dva} 
  N.~Yunes and X.~Siemens,
  Gravitational wave tests of general relativity with ground-based detectors and pulsar timing-arrays,
  Living Rev.\ Relativity {\bf 16}, 9 (2013)
  [arXiv:1304.3473 [gr-qc]].
  
\bibitem{Riess:1998cb} 
  A.~G.~Riess {\it et al.} (Supernova Search Team Collaboration),
  Observational evidence from supernovae for an accelerating universe and a cosmological constant,
  Astron.\ J.\  {\bf 116}, 1009 (1998)
  [astro-ph/9805201].

\bibitem{Perlmutter:1998np} 
  S.~Perlmutter {\it et al.}  (Supernova Cosmology Project Collaboration),
  Measurements of Omega and Lambda from 42 high redshift supernovae,
  Astrophys.\ J.\  {\bf 517}, 565 (1999)
  [astro-ph/9812133].
  
\bibitem{Clifton:2011jh} 
  T.~Clifton, P.~G.~Ferreira, A.~Padilla and C.~Skordis,
  Modified Gravity and Cosmology,
  Phys.\ Rept.\  {\bf 513}, 1 (2012)
  [arXiv:1106.2476 [astro-ph.CO]].
 
\bibitem{Hassan:2011zd} 
  S.~F.~Hassan and R.~A.~Rosen, 
  Bimetric Gravity from Ghost-free Massive Gravity,
  J.\ High Energy Phys.\ {\bf 02}, (2012) 126. 

\bibitem{deRham:2010ik} 
  C.~de Rham and G.~Gabadadze, 
  Generalization of the Fierz-Pauli Action,
  Phys.\ Rev.\ \textbf{D82}, 044020 (2010)
  [arXiv:1007.0443 [hep-th]].

\bibitem{deRham:2010kj} 
  C.~de Rham, G.~Gabadadze and A.~J.~Tolley,
  Resummation of Massive Gravity,
  Phys.\ Rev.\ Lett.\  {\bf 106}, 231101 (2011)
  [arXiv:1011.1232 [hep-th]].

\bibitem{Hassan:2011hr} 
  S.~F.~Hassan and R.~A.~Rosen, 
  Resolving the Ghost Problem in non-Linear Massive Gravity,
  Phys.\ Rev.\ Lett.\ \textbf{108}, 041101 (2012). 
  [[arXiv:1106.3344]].  

\bibitem{Hinterbichler:2011tt} 
  K.~Hinterbichler,
  Theoretical Aspects of Massive Gravity,
  Rev.\ Mod.\ Phys.\  {\bf 84}, 671 (2012)
  [arXiv:1105.3735 [hep-th]].
        
\bibitem{deRham:2014zqa} 
  C.~de Rham,
  Massive Gravity,
  Living Rev.\ Relativity {\bf 17}, 7 (2014)
  arXiv:1401.4173 [hep-th].
  
\bibitem{D'Amico:2011jj} 
  G.~D'Amico, C.~de Rham, S.~Dubovsky, G.~Gabadadze, D.~Pirtskhalava and A.~J.~Tolley,
  Massive Cosmologies,
  Phys.\ Rev.\ D {\bf 84}, 124046 (2011)
  [arXiv:1108.5231 [hep-th]].
  
\bibitem{DeFelice:2012mx} 
  A.~De Felice, A.~E.~Gumrukcuoglu and S.~Mukohyama,
  Massive Gravity: Nonlinear Instability of the Homogeneous and Isotropic Universe,
  Phys.\ Rev.\ Lett.\  {\bf 109}, 171101 (2012)
  [arXiv:1206.2080 [hep-th]].
  
\bibitem{Gumrukcuoglu:2012aa} 
  A.~E.~Gumrukcuoglu, C.~Lin and S.~Mukohyama,
  Anisotropic Friedmann-Robertson-Walker universe from nonlinear massive gravity,
  Phys.\ Lett.\ B {\bf 717}, 295 (2012)
  [arXiv:1206.2723 [hep-th]].

\bibitem{D'Amico:2012zv} 
  G.~D'Amico, G.~Gabadadze, L.~Hui and D.~Pirtskhalava,
  Quasidilaton: Theory and cosmology,
  Phys.\ Rev.\ D {\bf 87}, 064037 (2013)
  [arXiv:1206.4253 [hep-th]].

\bibitem{Izumi:2013poa} 
  K.~Izumi and Y.~C.~Ong,
  An analysis of characteristics in nonlinear massive gravity,
  Classical Quantum Gravity {\bf 30}, 184008 (2013)
  [arXiv:1304.0211 [hep-th]].

\bibitem{Deser:2013eua} 
  S.~Deser, K.~Izumi, Y.~C.~Ong and A.~Waldron,
  Massive gravity acausality redux,
  Phys.\ Lett.\ B {\bf 726}, 544 (2013)
  [arXiv:1306.5457 [hep-th]].

 \bibitem{Comelli:2011zm}  
  D.~Comelli, M.~Crisostomi, F.~Nesti and L.~Pilo, 
  FRW cosmology in ghost free massive Gravity from bigravity,
  J.\ High Energy Phys.\ {\bf 03} (2012) 067; {\bf 06} (2012) 020. 

 \bibitem{Yamashita:2014cra} 
  Y.~Yamashita and T.~Tanaka,
  Mapping the ghost-free bigravity into braneworld setup,
  J.\ Cosmolo.\ Astropart.\ Phys.\ {\bf 06} (2014) 004
  [arXiv:1401.4336 [hep-th]].

\bibitem{Dvali:2000hr} 
  G.~R.~Dvali, G.~Gabadadze and M.~Porrati,
  4-D gravity on a brane in 5-D Minkowski space,
  Phys.\ Lett.\ B {\bf 485}, 208 (2000)
  [hep-th/0005016].
 
 \bibitem{DeFelice:2013nba} 
  A.~De Felice, T.~Nakamura and T.~Tanaka,
  Possible existence of viable models of bi-gravity with detectable graviton oscillations by gravitational wave detectors,
  Prog. Theor. Exp. Phys. 2014, 43E01 (2014)
  [arXiv:1304.3920 [gr-qc]].

\bibitem{Will:1997bb} 
  C.~M.~Will, 
  Bounding the mass of the graviton using gravitational wave observations of inspiralling compact binaries,
  Phys.\ Rev.\ D \textbf{57}, 2061 (1998). 
  [gr-qc/9709011].  

\bibitem{Finn:2001qi} 
  L.~S.~Finn and P.~J.~Sutton,
  Bounding the mass of the graviton using binary pulsar observations,
  Phys.\ Rev.\ D {\bf 65}, 044022 (2002)
  [gr-qc/0109049].
  
\bibitem{Yagi:2009zm} 
  K.~Yagi and T.~Tanaka, 
  Constraining alternative theories of gravity by gravitational waves from precessing eccentric compact binaries with LISA,
  Phys.\ Rev.\ D \textbf{81}, 064008 (2010); \textbf{81}, 109902(E) (2010)
 [arXiv:0906.4269 [gr-qc]].  

\bibitem{Hazboun:2013pea} 
  J.~S.~Hazboun and S.~L.~Larson,
  Limiting alternative theories of gravity using gravitational wave observations across the spectrum,
  arXiv:1311.3153 [gr-qc].

\bibitem{Fierz:1939ix} 
  M.~Fierz and W.~Pauli,
  On relativistic wave equations for particles of arbitrary spin in an electromagnetic field,
  Proc.\ Roy.\ Soc.\ Lond.\ A {\bf 173}, 211 (1939).
  
  
\bibitem{Cutler:1994ys} 
  C.~Cutler and E.~E.~Flanagan,
  Gravitational waves from merging compact binaries: How accurately can one extract the binary's parameters from the inspiral wave form?,
  Phys.\ Rev.\ D {\bf 49}, 2658 (1994)
  [gr-qc/9402014].
  
\bibitem{Poisson:1995ef} 
  E.~Poisson and C.~M.~Will,
  Gravitational waves from inspiraling compact binaries: Parameter estimation using second postNewtonian wave forms,
  Phys.\ Rev.\ D {\bf 52}, 848 (1995)
  [gr-qc/9502040].

\bibitem{Barvinsky:2002gg} 
  A.~O.~Barvinsky, A.~Y.~.Kamenshchik, C.~Kiefer and A.~Rathke,
  Radion induced graviton oscillations in the two brane world, 
  Phys.\ Lett.\ B, {\bf 571}, 229 (2003).
  [hep-th/0212015].

\bibitem{Berezhiani:2007zf} 
  Z.~Berezhiani, D.~Comelli, F.~Nesti and L.~Pilo,
  Spontaneous Lorentz Breaking and Massive Gravity,
  Phys.\ Rev.\ Lett.\  {\bf 99}, 131101 (2007)
  [hep-th/0703264 [HEP-TH]].
  
\bibitem{DelPozzo:2011pg} 
  W.~Del Pozzo, J.~Veitch and A.~Vecchio,
  Testing General Relativity using Bayesian model selection: Applications to observations of gravitational waves from compact binary systems,
  Phys.\ Rev.\ D {\bf 83}, 082002 (2011)
  [arXiv:1101.1391 [gr-qc]].
    
\bibitem{Vallisneri:2012qq} 
  M.~Vallisneri,
  Testing general relativity with gravitational waves: A reality check,
  Phys.\ Rev.\ D {\bf 86}, 082001 (2012)
  [arXiv:1207.4759 [gr-qc]].
      
\bibitem{DelPozzo:2014cla} 
  W.~Del Pozzo, K.~Grover, I.~Mandel and A.~Vecchio,
  Testing general relativity with compact coalescing binaries: comparing exact and predictive methods to compute the Bayes factor,
  Classical Quantum Gravity {\bf 31}, 205006 (2014)
  arXiv:1408.2356 [gr-qc].
  
\bibitem{AdvLIGOZDHP}
  https://dcc.ligo.org/LIGO-T0900288/public.

\bibitem{Boulware:1973my} 
  D.~G.~Boulware and S.~Deser, 
  Can gravitation have a finite range?,
  Phys.\ Rev.\ D \textbf{6}, 3368 (1972).
  
\bibitem{Comelli:2012db}  
  D.~Comelli, M.~Crisostomi and L.~Pilo,
  Perturbations in massive gravity cosmology,  
  J.\ High Energy Phys.\ \textbf{06} (2012) 085. 
   
\bibitem{Vainshtein:1972sx} 
  A.~I.~Vainshtein,
  To the problem of nonvanishing gravitation mass,
  Phys.\ Lett.\ B {\bf 39}, 393 (1972).

\bibitem{Apostolatos:1995pj} 
  T.~A.~Apostolatos,
  Search templates for gravitational waves from precessing, inspiraling binaries,
  Phys.\ Rev.\ D {\bf 52}, 605 (1995).
  
\bibitem{Cornish:2011ys} 
  N.~Cornish, L.~Sampson, N.~Yunes and F.~Pretorius,
  Gravitational Wave Tests of General Relativity with the Parameterized Post-Einsteinian Framework,
  Phys.\ Rev.\ D {\bf 84}, 062003 (2011)
  [arXiv:1105.2088 [gr-qc]].
  
\bibitem{Punturo:2010} 
  M.~Punturo {\it et al}.,
  The Einstein Telescope: a third-generation gravitational wave observatory,
  Classical Quantum Gravity {\bf 27}, 194002 (2010).

 \bibitem{AmaroSeoane:2012je} 
  P.~Amaro-Seoane, S.~Aoudia, S.~Babak, P.~Binetruy, E.~Berti, A.~Bohe, C.~Caprini and M.~Colpi {\it et al.},
  Low-frequency gravitational-wave science with eLISA/NGO,
  Classical Quantum Gravity {\bf 29}, 124016 (2012)
  [arXiv:1202.0839 [gr-qc]].

\bibitem{Seto:2001qf} 
  N.~Seto, S.~Kawamura and T.~Nakamura,
  Possibility of Direct Measurement of the Acceleration of the Universe using 0.1-Hz Band Laser Interferometer Gravitational Wave Antenna in Space,
  Phys.\ Rev.\ Lett.\  {\bf 87}, 221103 (2001)
  [astro-ph/0108011].

\bibitem{Kawamura:2006up} 
  S.~Kawamura, T.~Nakamura, M.~Ando, N.~Seto, K.~Tsubono, K.~Numata, R.~Takahashi and S.~Nagano {\it et al.},
  The Japanese space gravitational wave antenna DECIGO,
  Classical Quantum Gravity  {\bf 23}, S125 (2006).
    
 \bibitem{Kawamura:2011zz} 
  S.~Kawamura, M.~Ando, N.~Seto, S.~Sato, T.~Nakamura, K.~Tsubono, N.~Kanda and T.~Tanaka {\it et al.},
  The Japanese space gravitational wave antenna: DECIGO,
  Classical Quantum Gravity {\bf 28}, 094011 (2011).    

\end{thebibliography}
\end{document}